\documentclass[12pt]{article}
\usepackage{graphicx}
\usepackage{amssymb} 
\usepackage{amsmath}    
\usepackage{times}

\title{\vspace*{-30pt}Rhythms of Memory and Bits on Edge:\\ Symbol Recognition as a Physical Phenomenon}
\author{John M. Myers* and F. Hadi Madjid** \vspace*{12pt}\\ 
\small{*Harvard School of Engineering and Applied 
Sciences,}\vspace*{-2pt}\\ \small{Cambridge, MA 02138, USA;} \\
\small{**82 Powers Road, Concord, MA 01742, USA}}

\begin{document}
\maketitle

\begin{abstract}
Preoccupied with measurement, physics has neglected the need, before anything can be measured, to \emph{recognize} what it is that is to be measured. The recognition of symbols employs a known physical mechanism. The elemental mechanism---a damped inverted pendulum joined by a driven adjustable pendulum (in effect a clock)---both recognizes a binary distinction and records a single bit.  Referred to by engineers as a ``clocked flip-flop,'' this paired-pendulum mechanism pervades scientific investigation. It shapes evidence by imposing discrete phases of allowable \emph{leeway} in clock readings; and it generates a mathematical form of evidence that neither assumes a geometry nor assumes quantum states, and so separates \emph{statements of evidence} from further assumptions required to \emph{explain} that evidence, whether the explanations are made in quantum terms or in terms of general relativity.  Cleansed of unnecessary assumptions, these expressions of evidence form a platform on which to consider the working together of general relativity and quantum theory as explanatory language for evidence from clock networks, such as the Global Positioning System.  Quantum theory puts Planck's constant into explanations of the required timing leeway, while explanations of leeway also draw on the theory of general relativity, prompting the question: does Planck's constant in the timing leeway put the long known tension between quantum theory and general relativity in a new light?
\vspace{20pt}
 \end{abstract}

\section{Introduction and Overview}\label{sec:1}

Over the years we watched ourselves working back and forth between writing
equations for clocks and signals on a blackboard and working with lasers,
lenses, and electronics on a work bench.  In the course of this experience we noticed the
role in physics of memories, both the memories of the investigators and the
memories of the digital computers they employ, and our eyes opened to
unsuspected vistas.  We speak of \emph{memory} as belonging to a \emph{party},
which can be a person, a computing machine, \emph{etc}.  As we mean it, a memory
is a device in which symbols are recorded and manipulated.  By \emph{memory} we
mean no static photograph, but a dynamic device in which the symbols recorded
can undergo changes from moment to moment. By \emph{symbols} we mean what is
recorded in a memory of a party, distinct from whatever propagates
externally from one party to another party, which we call a \emph{signal}.  The
elemental symbol carries a binary distinction: the bit.

What we call a party or a symbol or a signal depends on the level of
description, which can be finer or coarser. By a change in level of description,
what is termed ``a memory'' belonging to a single party can become several
memories belonging to distinct parties, with communications among them, and
\emph{vice versa}.  Thus the distinction between \emph{symbol} and \emph{signal}
is relative to the memory of a party, and both the memory and the party are
relative to a level of description.  As noted in Sec.~\ref{subsec:5.1} changes
in levels of descriptions will be seen to correspond to morphisms of graphs.

Regardless of how one imagines mathematical entities, their expression in
symbols is physical, \emph{e.g.} as ink on paper or voltages in a computer
memory.  Symbols in formulas and symbols of evidence from experiments live in
memories.  Thinking of symbols as physical attributes of memory, with associated
dynamics and rhythms, offers a physical analog of G\"odel coding: one can
inquire into the timing and location of symbols, both symbols of theory
expressing classical or quantum states and also symbols expressing evidence
extracted from experiments.

A familiar ``blackboard" picture of memory is the Turing-machine tape, divided
into squares; on each square a symbol ``0'' or a symbol ``1'' can be written or
erased.  Now lift up this abstraction to recall that the physical mechanism of
computer memory is a single device---what engineers call a clocked set-reset
flip-flop\ \cite{booth} that recognizes a binary symbol carried by a signal, and
as part and parcel of the act of recognition also acts as a memory device by
recording the symbol.  The flip-flop works as a damped inverted pendulum, a
hinge if you will, with its exposure to signals from outside cycled by a driven
adjustable pendulum, in effect a clock.  Noticing that a physical implementation
of a Turing machine depends on the flip-flop allows one to see symbols as
physical objects.  Then one can inquire into the motion of symbols, and into the
relation of that motion to concepts of spatial and temporal order.  The
paired-pendulum mechanism acts as a physical unit of computation and also, through
its participation in the machinery of radar, as a physical unit of geometry.

We make a distinction between \emph{recognizing a symbol in a signal} and
\emph{measuring the signal}.  In recognition, the hinge in the memory of a party
falls one way or the other to express the symbol; further, the
hinge-position-as-symbol can be copied to flip-flops in the memories of other
parties.  In contrast, measurement is idiosyncratic, characterized by error
bars, and no two instances of a measurement can be expected to agree exactly.
The results of a measurement, though
idiosyncratic, can be expressed in symbols (digitized), but only after waiting
for hinges to fall one way or the other, and with a (usually small) risk of
confusion.

\begin{figure}[b]
\centerline{\includegraphics[width=4.275in]{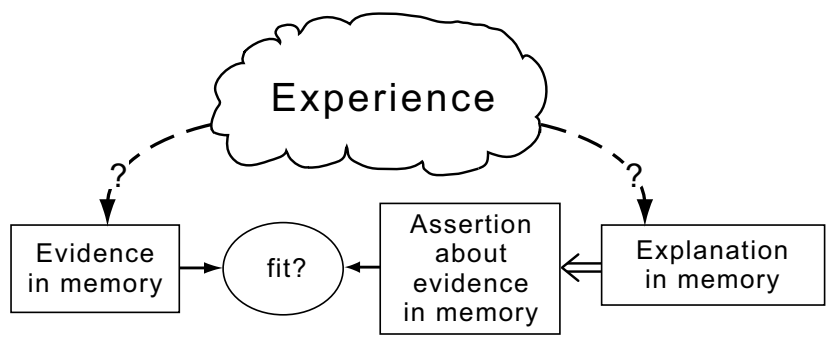}}
\caption{Evidence, mathematically expressed, compared with assertion implied by explanation.}\label{fig:1}
\end{figure}

Out of the buzzing world of experience, fingers on knobs, tweaking adjustments
to bring optics into alignment \emph{etc}., comes, one way or another,
\emph{evidence} from an experiment.  By \emph{evidence} we mean expressions in
mathematical language taken as reflecting experience on the work bench.  Theory,
quantum or otherwise, offers explanations, such as explanations in terms of
quantum state vectors and operators or explanations in terms of a
general-relativistic 4-manifold. An explanation asserts (rightly wrongly)
properties of evidence.  Experience evades direct comparison with theory,
but in memories symbols for evidence reflecting experience can be compared
against assertions about evidence implied by explanations. (See Fig.~\ref{fig:1}.)

\vfil

\eject
Recognizing the role of memory as the holder of evidence written in symbols
splits the question of the relation between theory and experience into two
questions:
\begin{enumerate}
\item How well does evidence in a memory reflect the experience of an
  investigator?\vadjust{\kern-6pt}
\item How well does an assertion about evidence implied by an
explanation fit actual evidence?
\end{enumerate}
Mathematical structures (\emph{e.g.} axioms) for explanations have been much
studied.  We raise the parallel question: what mathematical structures are to be
found or invented to express evidence?  In this report we concentrate on
structures of evidence recordable in the memories of communicating parties, to
do with the timing (not the content) of their communications.

One party communicates a symbol from its memory to the memory of another party
via a signal.  In propagating from one party to another, a signal deforms
unpredictably, so the recognition of a symbol carried by a signal must be
insensitive to a range of deformations. The damped inverted pendulum of the
paired-pendulum recognition mechanism offers this insensitivity, provided that the
\emph{rhythm of communication meshes the arrival of the part of a signal that
  carries a symbol with the phase of symbol recognition}.  The receiver must
look at the signal when the symbol is present, within some leeway but not too
much earlier or too much later.  By its dependence on the meshing of the part of
a signal that carries a symbol with a receiving party's phase of recognition,
the paired-pendulum mechanism of the flip-flop shapes evidence recordable from a
communications network by imposing discrete phases of the adjustable pendulum
for signal reception, leading to a single form of evidence, regardless of
whether explanations for evidence are stated in quantum terms or in terms of
general relativity.
 
The symbol recognized in a signal cannot be a function of the signal alone.  To
communicate, two parties must share some axioms in common, and also \emph{share
  a rhythm} that meshes their clocks with the signal propagating from one to the
other.  The rhythm, once acquired, must be maintained, and its maintenance
depends on reaching beyond the logic of symbol recognition: the rhythm of
symbol exchange is maintained not by recognitions but by \emph{measurements of
  signal arrivals relative to pendulum phases}.  These measurements are subject
to idiosyncrasies of each party, on which the other party must rely: an
\emph{intimacy} necessary to the communication of a symbol from one party to
another.

Radar as the instrument by which spacetime is conceived will be shown to have an
analog in the timing of symbols communicated among memories of a synchronized
network. (Indeed a working radar depends on the communication of symbols, such
as those identifying targets.)  Evidence of the timing of signals transmitted
and received in a network of communicating parties, recorded in their memories,
has a mathematical form independent of metric assumptions involved in
explanations based on the special or general theory of relativity.  We show this
form in a record format which we relate functorially to colored, directed
graphs.

The graphs expressing records of communication networks, such the Global
Positioning System (GPS), assume neither a general-relativistic geometry, nor
quantum states.  Because of this freedom from additional assumptions needed for
one or the other form of explanation, we will show how the graphical expression
of evidence offers a platform on which to negotiate the joint participation of
quantum theory and general relativity in explanations of evidence from networks
of communicating parties.

\section{Mechanism that Recognizes and Records}\label{sec:2}

We think of a memory as belonging to a communicating \emph{party}, a person or a
machine.  As a first cut, model a \emph{party} by a Turing machine moved by a
\emph{driven adjustable pendulum}---a clock with a faster-slower
lever. Following Turing\ \cite{turing}, we think of the history of a party's
memory as segmented into moments interspersed by moves, but, unlike Turing's
history of a memory as a sequence of snap shots at successive moments, we need
to inquire into what happens during a move in which the symbols in memory can
change.  Thus we view Turing's ``move'' not as something structureless but as a
phase of positive duration, during which there can be measurements of clock
readings. Picture the clock that moves a Turing machine as moving its hand
cyclically around a circle marked in subdivisions of the unit interval, so that
a reading of the hand position is the clock reading modulo integers.  Take the
phase `move' to be an interval of the circle that includes the position ``12
o'clock'' at the top of dial and the phase `moment' as a disjoint interval that
includes the ``6 o'clock'' position at the bottom of the dial.

At the level of description appropriate to an engineer who probes the operation
of a computer memory, the memory itself becomes a network of communicating
``sub-parties,'' each with a piece of the computer memory.  Symbols are conveyed
by signals from one piece of memory to another.  Because of uncontrolled
deformations as the signal propagates and because, on the workbench, no two
things ever get built quite alike, the signal that carries a symbol to a
receiving sub-party is subject to unpredictable deformations; and beyond these
practicalities, lower limits to signal variability are implied by quantum
theory.  For this reason, the mechanism for recognizing a symbol carried by a
signal must be made insensitive to small variations in the signal; \emph{i.e.}, the
signal has to be allowed a certain \emph{leeway} in both its shape and its
timing.  In terms of differential equations, recognizing a single symbol
regardless of a certain variation in the signal requires an attractor leading to
each symbol, with the implication that between attractors there are unstable
equilibria. The insensitivity to variations in the signal requires damping,
in conflict with any quantum explanation that invokes only the unitary evolution
of a Schr\"odinger equation.

Physically, the simplest memory element for recording a choice between two
symbols consists of paired pendulums, one inverted and damped, with two stable
positions and an unstable equilibrium between them, the other the adjustable
pendulum of a clock, swinging through phases as part of a rhythm of
communication, opening and shutting a gate to allow a signal to flip over or not
to flip over the inverted pendulum that holds a bit. A ``bit'' is thus a snap
shot of a livelier creature---a recognition-and-memory device that not only can
display a ``0'' or a ``1'' but, when the rhythm of its operation is disturbed,
can teeter in an unstable equilibrium like a flipped coin landing on edge, where
it can hang, lingering, with no sharp limit on how long it can take to show a
clear head or tail.  We are to think of a bit not as a 0 or 1 on a Turing tape
but as the position of the inverted pendulum at a moment \footnote{Attending to
  the dynamics of writing and reading a 1-bit record retrieves a critical
  element abstracted out of sight by Turing's ``machine'' and Shannon's
  channel\ \cite{shannon48}.}. In computer hardware, the inverted pendulum gated
by a clock is called a clocked set-reset (S-R) flip-flop\ \cite{booth}.

Without adequate maintenance of the rhythm of communication the part of a signal
in which a bit is to be recognized can arrive at a receiving party in a race
with the closing of the gate, resulting in ``runt signal'' squeaking through the
gate, big enough to push the inverted pendulum (think of a hinge) part way
but not all the way over, leaving the hinge hung up in an unstable ``in between''
state\ \cite{glitch,glitch2,glitch3}.  We say the signal straddles a timing boundary.

\subsection{Fan-out}\label{subsec:2.1}
Known to engineers concerned with the synchronization of digital communications,
such hang-up causes logical confusion.  Computation requires acts of copying
symbols: a symbol in flip-flop A at one moment is copied into two flip-flops,
say B and C, at a later moment, so that whatever bit value was in A at the
earlier moment appears in both B and C at the later moment---both hold 0 or both
hold 1; one speaks of ``fan-out.''  If flip-flop A hangs up in an unstable
equilibrium, then flip-flops B and C not only may hang up, but can ``fall
differently'' so that the symbol in B, instead of matching that in C, conflicts
with it.

In revealing conflicts in response to an unstable condition of a flip-flop A,
the fan-out from A to flip-flops B and C also offers a means of detecting
unstable conditions, which has been used to show a roughly exponential decline
with waiting time of the probability of disagreement between B and C, resulting
in the measurement of a \emph{half-life} $\tau$ of the instability\ \cite{aop05}.
(For silicon integrated circuits we found $\tau$ to be close to 1 ns.  Modern
gallium arsenide circuits operate much faster, and efforts to shorten their
half-life are underway, but so far their ratio of half-life to cycle period is
not much less than that for silicon \cite{GaAs}.)
In quantum explanations, one describes the inverted pendulum by a wave function,
putting Planck's constant into the relation between the short time constant
required for rhythmic operation of the flip-flop and the long time that must be
waited for it to settle down when subject to the straddling of timing boundaries
and the ensuing runt pulses \cite{aop05}.\looseness=-1

In its use to decide a race among more than two signals, the teetering hinge of
a flip-flop has a noteworthy consequence.  Consider the case of a three-way race
among signals $A$, $B$, and $C$ arriving at a clock.  A world line in a
general-relativistic explanation of this clock corresponds on the workbench not
to one device but to several interconnected devices.  Each of the three signals
fans out to allow three separate pairwise comparisons of ``which came before
which''.  In a close race, teetering in all three pairwise comparisons can
result in finding: $A < B < C$, and $C < A$, violating the transitivity of an
ordering relation, and suggesting a limit on the validity of even local temporal
ordering.  Making sense out of temporal order requires distinguishing the
question of which cycle a symbol recognition occurred from the question of when
within a cycle did a signal arrive.\looseness=-1
\vspace*{8pt}

\noindent Remarks:\vadjust{\kern-6pt}
\begin{enumerate}
\item To reduce the risk of disagreement between B and C, it suffices to wait
  after the setting of A to the reading of B and C.  The literature on digital
  circuits discusses the related use of ``arbiters''---of which there are two
  types, one that might take forever, the other that might generate confusion.\vadjust{\kern-6pt}
\item Weeks after a given day, GPS publishes corrections to coordinates for events
  that it issued on that day, derived from subsequent cross comparisons among
  its clock readings recorded at the transmission and reception of radio
  signals. Although the process of comparing and correcting may yet be greatly
  speeded, not only does the delay in communicating comparisons limit how
  quickly one can determine what the clock readings ``should have been,'' but an
  additional delay is imposed by the balancing instrument used to convert analog
  measurements to digital signals suitable for communication.
\end{enumerate}

\section{Idiosyncratic Maintenance of Shared Rhythms}\label{sec:3} 

For theoretical purposes, we assume the conceptually simplest (but not the most used) scheme
for digital communications, called \emph{synchronous communication} \cite{meyrI},
which offers the fastest response.  In synchronous communication a receiver
rec\-ognizes symbols one by one (without use of sample-and-hold techniques
\cite{meyrII}).  Synchronous communication from a party A to a party B, moved by
clocks A and B, respectively, requires that a symbol be transmitted from clock A
while A's clock hand is in the 12 o'clock ``move'' phase and must arrive a B while B's
clock hand is also in a 12 o'clock ``move'' phase.  

Clocks, including the atomic clocks used to generate International Atomic Time
(TAI), drift unpredictably in rate, leading eventually to unbounded phase drift
between two nearby clocks, with the result that clocks function only in a
network of comparisons that guide adjustments of clock rates over some (possibly
small) range. In addition, communications involve other perturbing
circumstances, including Doppler shifts among parties in motion.
Unless the clock of a receiving party can be maintained so the phases of
reception are aligned with the arrivals of symbol-carrying signals, the
recognition of a symbol carried by a signal fails.  Suppose that the conditions
of phasing allowing synchronous communication between two parties have been
brought into being---a story in itself \cite{meyrII}.  To maintain these
conditions over a succession of symbols requires more or less continual
adjustment of the motion of the clocks: their accelerations, their rates of
ticking, or both.\footnote{One speaks of various phase-locked loops
  \cite{meyrI}.}  In all cases the adjustment is guided by departures from
nominal behavior of the arriving symbols relative to an imagined center of the
phase of reception, much as steering an automobile toward the center of a lane
depends on noticing and responding to its departure from the center.\looseness=-1

How then to determine the departures in the clock reading of a receiving party
at a signal arrival?  Let the reading of the clock of the a receiving party
relative to the 12 o'clock center of the receptive move, modulo integers, be
symbolized by $\Delta$.  In order to guide adjustment necessary to the
maintenance of synchronous communication , the offset symbolized by $\Delta$ has
to be made to act on a ``lever'' (as in the lever on the back of a wind-up clock
by which its rate of ticking is adjusted).  (If $\Delta >0 $ the receiver clock
needs to be showed down relative to the arriving symbols, and speeded if $\Delta
< 0$.)

In hardware, the symbol ``$\Delta$'' never appears.  For example, one way to guide
adjustment is by ``bang-bang'' control that responds to whether the part of a
signal that carries a signal arrives before or after a nominal clock reading
within the receptive phase. For this a logical AND gate is used not as part of a
device for recognizing symbols but as a measuring device in a feedback loop that
controls the rate of ticking of the receiving party. The AND gate is opened at
the beginning of the cycle to the arrival of the signal but turned off at the
nominal reading.  If the signal arrives well before the turn-off it passes
through the AND gate to put a pulse of charge on a capacitor.  A running average
of the charge on the capacitor controls the faster-slower lever of the party's
clock.  Close races between the arriving signal and the turn-off produce runt
pulses without causing any logical confusion, for the runt pulses never need to
be recognized as symbols; instead the pulses, runt or not, pile up like gravel
that is shoveled without the stones being counted.

The point is that the fine-grained determination of clock reading
within a receptive phase at the arrival of a symbol cannot be \emph{recognized}
as a symbol but requires something distinctly different, which we call
\emph{measurement}, as follows.  Symbol recognition depends not only on leeway
but also on avoiding ``straddling of boundaries.''  To recognize a symbol, such
as the arrival of a pawn on a square of a chess board, the act of looking must
be coordinated with the arrival so that a party looks while the pawn is in the
square and not sliding over a boundary that it straddles as it moves.  
It is these conditions of ``no straddle'' and ``leeway'' that allow two parties
to agree exactly in their recognitions of symbols.  In contrast to the
recognition of symbols, we speak of \emph{measurement} as in the determination
of a mass in a balance, for which no two parties can expect to agree exactly;
instead one speaks of error bars.  The idiosyncratic variations among parties
resulting in error bars are inescapable precisely because of the straddling of
boundaries and the lack of leeway.

Although distinct, `measuring' and `recognizing' depend on one another.  For
example, in measuring using a balance instrument, I have to recognize weight A,
weight B, and that ``they balance'' or ``the balance tips toward A,''  and
if I am wrong in such noticing, my measuring makes no sense.  Indeed, in spite
of their neglect in physics education, recognitions are essential to logic,
without which physics collapses.  Going the other way, recognitions basic to
logical communication turn out to take place in rhythms that require
maintenance---adjustments to clock rates---guided by measurements.

With the distinction between recognizing and measuring in mind, we return to the
issue of determining a departure from a desired clock hand position within a
phase at the arrival of (the center of) a symbol carried by a signal.  In its
use to recognize a symbol the mechanism of an inverted pendulum must be
insensitive to the very timing variations of interest within the leeway of the
phase of reception.  When finer-grained distinctions necessary to determining
the clock hand position are implemented, straddling of boundaries is
unavoidable, and the distinctions cannot be \emph{recognitions} but depend on
\emph{measurements}.  Altogether we arrive at:
\begin{quote}
  Fine-grained ``local clock readings''---necessary to maintaining rhy\-thms
  essential to the communication of logical symbols---constitute measurements,
  idiosyncratic in that no exact agreement can be expected between any two
  measuring parties.
\end{quote}

Only in special situations can a receiving party recognize in a signal the
symbol intended by a transmitting party. For example, when two people converse:
each person's ear hears the symbols what the other's mouth puts into a spoken signal.
For communication the two parties have to share not only concepts but a rhythm,
and the establishment of that rhythm requires reaching beyond logical
recognitions to rely on necessarily idiosyncratic measurements that guide the
adjustments needed to maintain the rhythm.  The conditions of shared concepts
and a shared rhythm necessary to communication can reasonably be called
\emph{intimacy}.  Without this intimacy of communication in which symbols are
conveyed, there can be no logic, mo mathematics, and no physics.

\section{The Distinct Forms of Evidence and Explanations}\label{sec:4}

In working back and forth between experiments on the optics bench and writing
quantum states on the blackboard, we saw lens holders and lenses and lasers
on the optics bench, but \emph{no} quantum state vectors.  But \emph{must} state
vectors be invisible on the bench?  

Nobody can lay formulas on top of an optics bench to see if they fit.  To be
compared with mathematically expressed explanations raw experience with lenses
and mirrors has to be first reflected into \emph{evidence written in symbols of a mathematical system based on axioms}, recorded in a memory.  So our question
became: can mathematically expressed evidence in a record ever determine its own
explanation?  The answer hinges on a striking property of quantum theory.

In pre-quantum physics, including general relativity, the
mathematical system available for expressing evidence involves the same
axioms as that for expressing explanations.  Quantum theory differs by invoking
two distinct mathematical systems, Hilbert-space constructions
for \emph{explanations}, and a distinct other system of probability
measures for assertions about \emph{evidence} implied by an explanation:
\begin{center}
  \begin{tabular}{cccc}
  \label{eq:born}
\large{$\text{tr}$}&\large{$[M(\omega)\rho]$} &\large{=} & \large{$\Pr(\omega)$}
\\
\textbf{map}&\textbf{explanation}& &\textbf{assertion}\\
&&&\textbf{about evidence}\\
\end{tabular}
\end{center}
where $\omega$ is an outcome recognized in a signal from the experiment, and the
repeatable preparation is symbolized by a positive operator-valued measure $M$
and a density operator $\rho$.  

The \emph{trace} maps explanations as Hilbert-space constructions to assertions
about evidence as probability measures. In spite of the mapping from
explanations to assertions about evidence, the separation of axioms for
Hilbert-space explanations from axioms for expressing probabilistic assertions
about evidence matters, because the mapping is \emph{not} injective.  Without
injectivity the inverse problem can have no unique answer: the evidence from
experiments can never fully determine its explanation in terms of the quantum
states and operators; rather, there is always freedom of choice for an
explanation of given experimental outcomes \cite{aop05,ams02,tyler07}, a choice
outside of logic, a \emph{guess} reminiscent of the choice in mathematics of a
model of an axiom system.  The scientist who makes the guess is part of the
story of science.

The separation of axioms for assertions about evidence from axioms for
explanations let us hope for an analogous separation in spacetime physics
between axioms to express evidence and additional axioms for whatever geometric
theory one chooses for explanations.  The separation developed below has the
potential to relieve confusion in the notion of a \emph{reference system} as
used in geodesy as an underpinning for a \emph{reference frame}
\cite{soffel03}.  The reference system stated in the International Astronomical
Union (IAU) resolutions of 2000, consists of a general-relativistic spacetime
along with coordinate charts and a metric tensor field \cite{soffel03}.  By
mixing in the general-relativistic geometry, this reference system assumes more
than is necessary to express evidence; moreover its expression of evidence neglects some significant experience with GPS devices.  By noticing the distinct phases
of any cycle of operation of a clock-driven memory we offer what appears to us
to be a substantial repair: a \emph{reference system} for evidence, separated
out from additional assumptions of a geometry in terms of which to
\emph{explain} that evidence.

\section{Clock Readings Recorded}\label{sec:5}

We arrive at a formal structure of evidence of the timing of communications (not
the content) recordable by communicating parties as follows.
\begin{enumerate}
\item First, imagine a \emph{party} (person or machine) as an implemented
  \emph{Turing machine} moved by a \emph{driven adjustable pendulum}---a clock
  with a faster-slower lever.\vadjust{\kern-6pt}
 \item Augment the Turing machine with a communication capability.  Each party is
  moved by its own driven adjustable pendulum (clock).  A party $A$ converts a
  symbol to a signal in which party $B$ recognizes the symbol. 
  (For a given machine, a symbol is internal to its memory, a signal is
  external.)\vadjust{\kern-6pt}
\item Assume that the driven adjustable pendulum of a party turns a clock hand
  one revolution per cycle of the swinging pendulum, so that we can speak of a
  \emph{phase} in which transmission or reception happening as the clock hand
  passes the 12 o'clock mark at the top of the dial.\vadjust{\kern-6pt}
\item At each passage of the clock hand of a party past the 6 o'clock mark (at
  the ``bottom of the dial'') a \emph{count of cycles} is incremented, giving a
  coarse measure, of temporal order, local to that party, within which the clock
  hand functions as a ``second hand'' to mark subdivisions. (Counting passages
  of the clock hand past the ``bottom of the dial'' assures that each 12 o'clock
  phase of a move belongs to a single cycle count rather than straddling
  adjacent counts; this counting involves recognition, not measurement.)\vadjust{\kern-6pt}
\item Each party measures the position of the clock hand on the dial as each
  symbol-bearing signal arrives within the receptive phase.\vadjust{\kern-6pt}
\item These (idiosyncratic) measurements guide rate
  adjustments to maintain signals arriving during the receptive phase.\vadjust{\kern-6pt}
\item All parties of a network record histories of\vadjust{\kern-8pt}
  \begin{enumerate}
  \item cycle count when they send or receive a signal,\vadjust{\kern-2pt} 
  \item (idiosyncratic) measurements of pendulum positions at symbol
    recognitions, and\vadjust{\kern-2pt}
\item pendulum rate adjustments. 
  \end{enumerate}
\item  Recorded histories can be communicated from one party to
  another. Assembled recorded histories are the form of
\emph{evidence of the timing of signals transmitted and received in a
network of communicating parties}.
  \end{enumerate}
Evidence of this form assumes neither the axioms of any geometry that might be
  chosen for \emph{explanations}, nor the axioms of quantum theory.

  \subsection{Functor from recorded histories to graphs}\label{subsec:5.1}

A history recorded by a party A maps to a fragment of a colored, directed graph,
as follows.
\begin{itemize}
\item Each count of cycles of A's clock maps to a vertex.\vadjust{\kern-6pt}
\item A directed edge runs from each vertex for a count of cycles to the vertex for
  the successor count.\vadjust{\kern-6pt}
\item A signal received by A at a cycle count $n$ is indicated by an additional
  vertex for the sending party and a signal edge from that vertex to the vertex of A
for cycle count $n$.\vadjust{\kern-6pt}
\item A signal transmitted by A at cycle count $n$ to another party is indicated
  by an additional vertex for the receiving party and a signal edge
from the vertex of party A at cycle count $n$ to the vertex for the receiving
party.
\end{itemize}

\noindent Coloring:
\begin{itemize}
\item The edge from a vertex to a successor vertex is colored by (a) party identity and
(b) the rate setting of the party.\vadjust{\kern-6pt}  
\item The edge for an incoming signal is colored
by (a) designation as a signal and (b) the fine-grained clock reading within the
receptive phase at the arrival of the symbol carried by the signal.\vadjust{\kern-6pt} 
\item  The edge
for an outgoing signal is colored by signal identity.\vadjust{\kern-6pt}
\item The vertex at the head of
an edge for a transmitted signal is colored by the identity of the receiving
party.\vadjust{\kern-6pt}
\item The vertex at the tail of an edge for a received signal is colored by
the identity of the transmitting party and by the cycle count (assumed to be
encoded in the signal) of the transmitting party at its move of transmission. 
\end{itemize}
    
\vspace{12pt}

\begin{figure}[h]
\begin{center}
\includegraphics[width=1.9in]{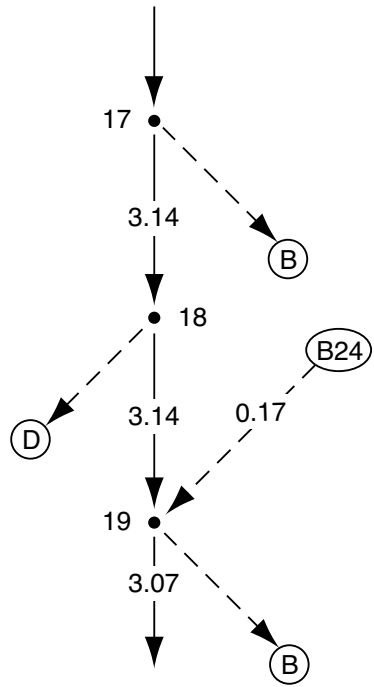}
\end{center}
\caption{Occurrence graph fragment for record of Party~A.}\label{fig:2} 
\end{figure}

For the record shown in Table 1, the corresponding graph fragments are shown in
Fig.~\ref{fig:2}.  As illustrated, a functor takes recorded histories to occurrence graphs, and in some cases to marked graphs and to Petri nets.

\begin{center}Table 1. History recorded in the memory of a party A\vadjust{\kern12pt}
\begin{tabular}{||ccccc||}\hline
Cycle & Event & Other:& Fraction & Cycle\\
count&&party&or rate&sent\\
\hline
\vdots &\vdots&\vdots&\vdots&\vdots\\\hline
17  & send &B  &  &  \\
&rate&&3.14&\\\hline
18&send&D&&\\
&rate&&3.14&\\\hline
19&rec'd&B&0.17&24\\
&rate&&3.07&\\
&send&B&&\\\hline
\vdots&\vdots&\vdots&\vdots&\vdots\\\hline
\end{tabular}
\end{center}  
\vfil

\begin{figure}[h]
\centerline{\includegraphics[width=5.2in]{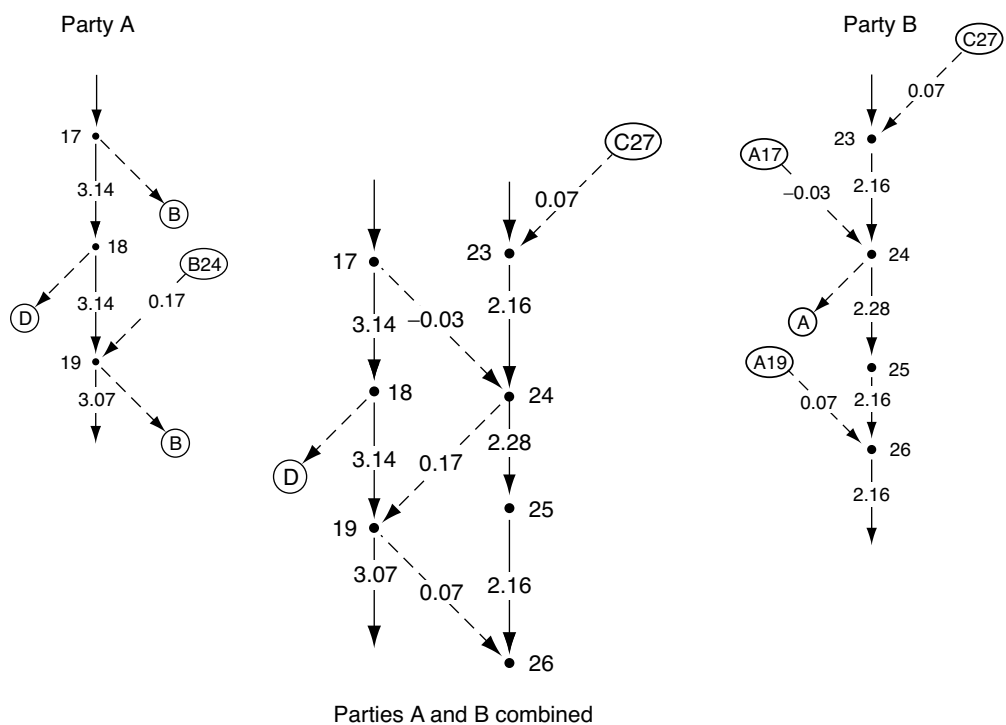}}
\caption{\ Graph fragments for record of Parties~A and~B combined.}\label{fig:3}
\end{figure}

\eject
Such graph fragments can be pasted together by condensing a signal edge of the
graph for party A for transmission to another party B and the signal edge for
reception of A's transmission by a party B into a single edge from the
transmission move of A to the reception move of B, as follows.  A vertex at the
head of a signal edge from a move of A at count $n$ is overlaid on a vertex
colored A$n$ at the tail of an edge to reception at a move of party B, the
vertex is removed and the signal arrows joined head to tail into a single
directed edge. This is illustrated in Fig.~\ref{fig:3}.

\begin{figure}[t]
\centerline{\includegraphics[width=2.4in]{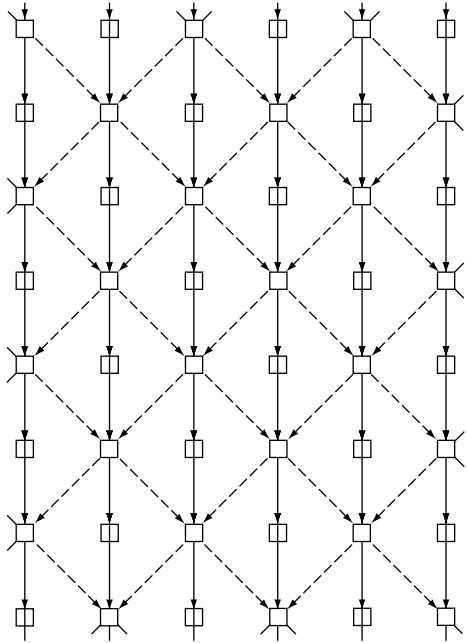}}
\caption{\ Fragment of occurrence graph for six parties exchanging signals with neighbors.}\label{fig:4}
\end{figure}

\begin{figure}[t]
\centerline{\includegraphics[width=2.75in]{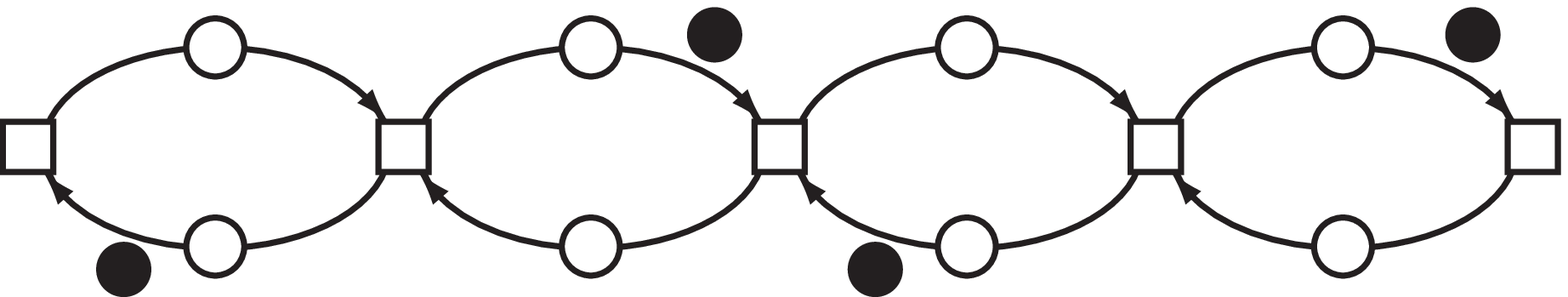}}
\caption{\ Marked graph obtained from folding occurrence graph for four parties ``around a cylinder."}\label{fig:5}
\vspace{25pt}
\centerline{\includegraphics[width=2.5in]{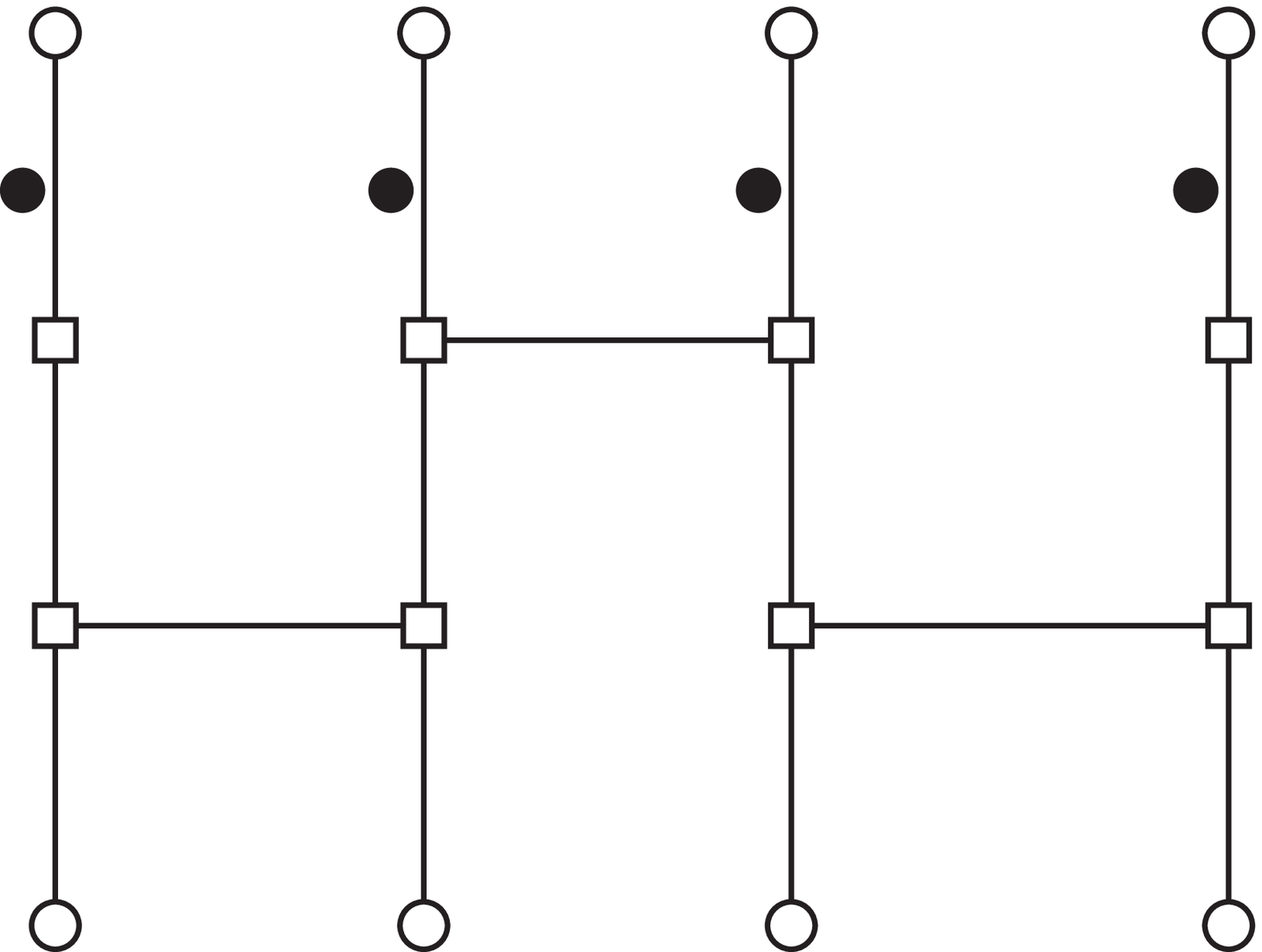}}
\caption{\ Marked graph as in Fig.~\ref{fig:5} redrawn as a role-activity graph, with a vertical trail for each of four parties.  A circle at the top of a trail is identified with the circle at the bottom of the trail, and vertices (square boxes) connected by a horizontal line are identified.  Vertical edges are understood to be downward-directed.}\label{fig:6}
\end{figure}

\begin{figure}[t]
\centerline{\includegraphics[width=3.18in]{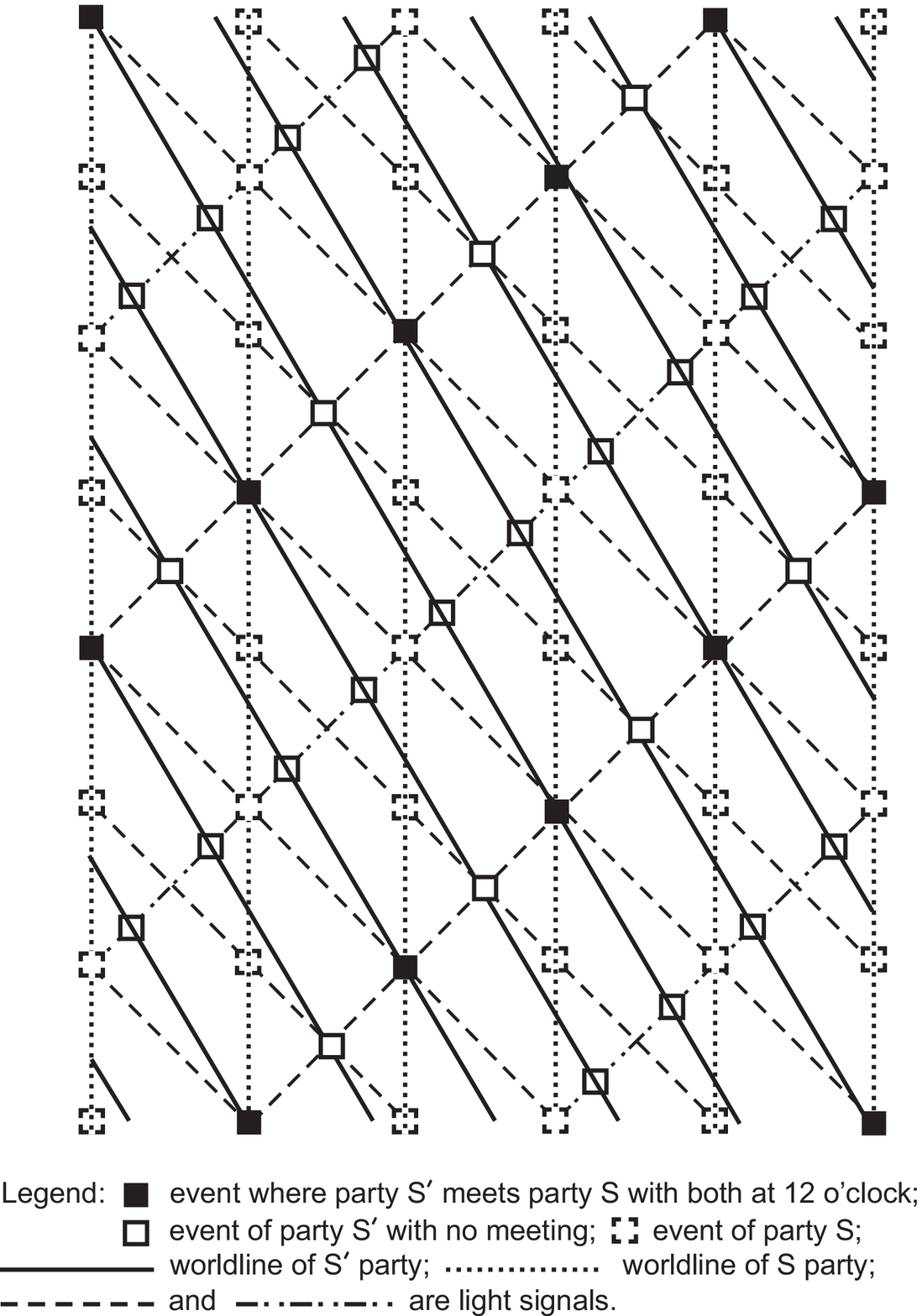}}
\caption{\ Occurrence graph for sets of parties S and S$^\prime$ moving past one another.  Solid boxes indicate a meeting between a party of one set and a party of another set.  All edges are directed downward.}\label{fig:7}
\end{figure}

Such graphs are essentially \emph{occurrence graphs} \cite{holtOcc},
specialized to exhibit a distinct trail for each party, with edges for signals
linking parties.  When ``analog'' measurements with their idiosyncrasies that
color the occurrence graphs are forgotten, the occurrence graph for a network
of communicating parties can exhibit symmetry, illustrated in Fig.~\ref{fig:4}.  In some
interesting cases, forgetting the coloring by fine-grained clock readings and
rate settings, an occurrence graph can be ``wrapped around'' to form a marked
graph \cite{mk_graph}, as in Figs.~\ref{fig:5} and~\ref{fig:6}.  Figure~\ref{fig:7} shows an example of an
occurrence graph for a network in which one set of parties is in motion relative
to another set of parties.

Occurrence graphs, marked graphs, and, more general Petri nets
\cite{peterson81,holt75} form categories with interesting graph morphisms.
Going the other way, one studies morphisms among network histories, aided by the
functor from network histories to occurrence graphs.

The graphs are objects of respective categories in which morphisms include (1)
isomorphisms from one induced subgraph to another; (2) inclusions; and (3)
epimorphisms in which certain stretches of clock image over several vertices for
moves along with neighbor-to-neighbor signals map to a single vertex.  Example:
view main memory as one party and view auxiliary memory as a second party; then
map the two parties into a single ``Turing-machine'' party.  

\begin{figure}[t]
\centerline{\includegraphics[width=3.25in]{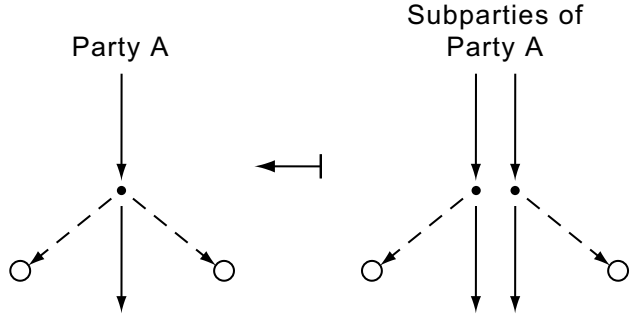}}
\caption{\ Condensation from finer level of detail.}\label{fig:8}
\vspace{8pt}\end{figure}

By another such map, illustrated in Fig.~\ref{fig:8}, a vertex at which two signal edges
meet a party can be seen as a condensation of a pattern involving two parties,
each with a vertex involving only one signal edge.  Occurrence graphs of this
form of ``no more than one signal per party vertex'' map to virtual braid
diagrams \cite{virtual}, and it will be interesting to see what interpretation,
if any, to make of virtual-braid isotopies.

A nice path to study more complex synchronization methods, including those used in GPS, employs the two-step procedure of choosing a sensible form for records of timing recordable in the memories of communicating parties, and then translating those records to graphs.
We expect different synchronization methods to produce different formats for
records, which in turn will imply different special properties of the occurrence
graphs to which they map.  For that reason the basic starting point is the
notion of the records recordable in the memories of communicating parties.

\subsection{Echo count}
A noteworthy property that can be defined by a network history and read from the
corresponding occurrence graphs is what we call
\emph{echo count}, which is an integer-valued measure relevant to
communications, defined to be the difference between the cycle count of a
transmitting clock A at the transmission to clock B and the cycle count at A at
which an echo from B is possible.  Let ec$(n,A.B.A)$ be the echo count for
transmission from A during cycle count $n$ to echo back from B to~A.
\ Note that:
\begin{enumerate}
\item The echo count can vary along a history in which one party receives a
  sequence of echoes from another party.\vadjust{\kern-6pt}
\item Except in special cases the echo count \emph{not} symmetric. For
  instance, clock of A can run twice as fast as clock of B, resulting in
  ec$(n,A.B.A)=$ 2ec$(n,B.A.B)$. 
\end{enumerate}

\section{General-Relativistic and Quantum Explanations}\label{sec:6}

So far we have concentrated on colored directed graphs as reference systems for
evidence.  Here we put in a word about explanations of such evidence, by looking
at what happens if one chooses to introduce the additional assumptions,
not required for expressing the evidence, but needed for explanations.

We start with general relativity. In order to explain synchronous communication of symbols from one clock to another in the language of general relativity we follow convention by modeling a clock as a smooth embedding $\gamma\colon t\mapsto \gamma(t)$ from a real interval $I\subset \mathbb{R}$ into 4-dimensional manifold $M$ with a smooth metric
tensor field $g$ of Lorentzian signature and time orientation, such that the
tangent vector $\dot{\gamma}(t)$ is everywhere timelike with respect to $g$ and
future-pointing \cite{perlick}.  

To express the positive duration of phases of moves, recall the distinction
between an embedding $\gamma$ as a curve---that is, a function from $I$ to $M$,
and the image of this curve as a 1-dimensional submanifold of $M$, denoted
$\text{image}\ \gamma$.  For lack of a better word, we call such an image a
\emph{thread}.  Think of a dial position attached to each point of the thread
for a party, and picture the thread for a clock that takes part in synchronous
communication as striped by the 12 o'clock phases in which transmission and
reception are allowable.

The form of a general relativistic explanation of evidence presented in the
colored occurrence graphs is then a corresponding network of threads, with
timelike threads for parties and lightlike threads for signals from thread to
thread.  Such a network of threads in a manifold with metric maps to an
assertion of evidence; however, as in the case of the trace as a map from
evidence to an assertion of evidence in quantum theory: the map from a network
of threads to a colored occurrence graph is \emph{not} injective: for given any
given colored occurrence graph displaying evidence, there is a freedom to change
the metric tensor and make a corresponding change in the convention for relating
physical clocks to proper clocks leaving unchanged the assertion of evidence
implied by the explanation.  Indeed, in applications such as GPS, one needs to
invoke non-gravitational forces, and these forces have to be estimated from
their effects on evidence, bringing in a much larger realm for free choice of
explanation.

\subsection{Paired computers}\label{subsec:6.1}

Consider a spacetime manifold and two parties $A$ and $B$ as non-intersecting,
threads colored by their respective clock readings. A change in clock rate is
expressed by a change in the coloring along the thread. If the manifold is flat,
it is always possible to adjust the clock rates in such a way that:
\begin{enumerate}
\item Synchronous communication can take place from $A$ to $B$ and from $B$ to
  $A$;\vadjust{\kern-6pt}
\item An event of $A$ can be chosen freely as a transmission event, provided the
  clock is reset, as represented by re-coloring the thread for $A$ so that the
  event corresponds to an integral clock reading;\vadjust{\kern-6pt}
\item Given a clock as thread $A$
colored by its reading, along with an integer $n_A$, there exists a clock
$B$ allowing for synchronous communication at echo distance ec$(n,A.B.A)=n_A$,
independent of cycle count $n$.
\end{enumerate}
The same holds in a curved spacetime if the clocks are not too far apart, which
is the case if for each event $p$ of the thread for A there be an event of the
thread for B within a radar neighborhood of $p$ with respect to the thread for
A, and \emph{vice versa} \cite{perlick}.

\subsection{Coordinated universal time}\label{subsec:6.2}

In 1967, the 13th General Conference on Weights and Measures specified the
International System (SI) unit of time, the second, in terms of a cesium atomic
clock rather than the motion of the Earth. Specifically, a second was defined as
the duration of 9,192,631,770 cycles of microwave light absorbed or emitted by
the hyperfine transition of cesium-133 atoms in their ground state, supposing
the atoms are undisturbed by external fields.  Two commercially available cesium
clocks functioning well can vary in rate by about 1 part in $10^{12}$, and
primary cesium standards approach 1 part in $10^{16}$.  Nonetheless, as clock
improve in their reproducibility, the size of discrepancies that matter keeps
shrinking.  For example the National Institutes of Science and Technology (NIST)
detects a rate shift between two optical clocks of $(4.1\pm 1.6)\times 10^{-17}$
when one clock is lifted against the Earth's gravity by 33 cm and this shift is
proposed as a basis for mapping the Earths gravitational field\cite{chou}.
Because of size of discrepancies that matter keeps shrinking in step with
improvements of precision, we continue to experience the circumstance that ``no
two clocks tick alike.''

For this reason the choice of $^{133}$Cs or any other clock design can only be a
partial specification of the clocks used to generate Coordinated Universal Time
(UTC). UTC actually makes use of a global system of signaling between clocks,
comparing clock readings at the arrival of signals, deciding what these readings
``would be'' if the clocks were proper clocks and the general relativistic
metric tensor were that assumed, and issuing ex-post corrections to readings of
clocks reported by national laboratories.  A big part in this inter-clock
communication is played by the Global Positioning System (GPS).Thus in practice,
a ``standard clock'' in not local to any single physical clock, but instead is a
creature of a network of communicating clocks governed by a scheme of
comparisons of signal arrivals that guide adjustments of clock rates, or, what
is the same in its effect on recorded times, corrections.  The second depends on
(a) a network of clock-driven communications, and (b) the assumption of general
relativity and of some particular choices of metric tensor field within that
theory.

By sorting out a reference system for evidence distinct from assumptions of
geometry---e.g. a choice of metric tensor field---we offer the opportunity to
put the choice-making aspect of UTC up on the table where it can be considered
more clearly, by virtue of a reference system for evidence independent of the
dynamical and indeed chaotic nature of the metric tensor field of general
relativity.

\subsection{Constraints on synchronization imposed by spacetime\\ curvature}\label{subsec:6.3}

It follows from Perlick's work \cite{perlick} that spacetime curvature imposes a
lower bound on the duration of phases of moves in a network of synchronous
communication by use of signals that propagate at the speed of light.  We note
however, as follows from the remarks above on paired clocks, that there is no
such bound if only two clocks constitute the network.  

The tightness of synchronization in a network of communicating parties is
indicated by the greatest required phase duration: the less the phase duration,
the tighter the synchronization.  The tightness of synchronization possible when
the network operation is restricted to a subregion is apt to be greater than for
the network over the region.  For this reason there can be no network that is
universally tightest over all subregions.  Applied to coordinate-generating
networks such as GPS, the implication is that for the highest precision over a
limited spacetime region, the scheme of synchronization must be specially
adapted to the limited region of interest.  For the future is will be
interesting to study the possibility of adapting clock networks to achieve the
tightest synchronization possible for particular uses in which a limited region
of spacetime is at issue.

\begin{figure}[h]
\centerline{\includegraphics[width=3in]{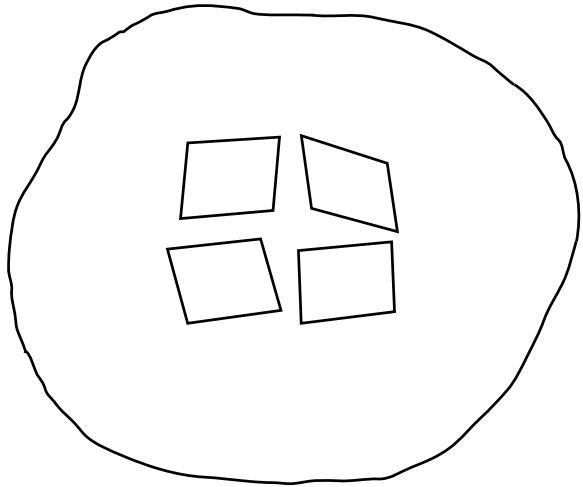}}
\caption{\ Square tiles laid on ``potatoid": the grout takes up the slack.}\label{fig:9}
\end{figure}

In a curved spacetime such as that appropriate to explain the Global Positioning
System, there are no Killing vector fields and indeed no exact isometries
linking two disjoint spacetime regions.  Yet there can be occurrence graphs for
a clock network that, once idiosyncratic clock readings and rate settings are forgotten, exhibit
exact isomorphisms from one graph fragment to another.  One can make an analogy
to isomorphisms among square tiles laid over a region of a ``potatoid,'' where variations
in the thickness of the grout take up the slack, as illustrated in Fig.~\ref{fig:9}. Such
isomorphisms come as close as one can get to resolving the need in quantum
theory to speak of repeated occurrences of the preparation of an experiment.

\vfil
\pagebreak

\subsection{Assertions of evidence implied by quantum explanations}\label{subsec:6.4}

Quantum mechanical explanations imply probability distributions for clock readings of a receiving party at the arrival of a signal within a receptive phase. (Indeed the distributions cannot be confined to a receptive phase, leading to occasional
logical failures that can be reduced in their disruptive effects by well-known
error-correction techniques, based on redundancy, but never reduced to the
vanishing point.)

But what to make of a probability measure for clock readings?  Experimentally,
one compares an asserted probability with relative frequencies of clock readings
at signal arrivals.  For this one has to identify many disjoint fragments of an
occurrence graph as pertaining to repetitions of a single quantum
``preparation.''  Assuming a flat spacetime, this identification perhaps
presents no problem.  In contrast, when one wants to work with quantum theory
adjoined to a curved spacetime of general relativity, the situation becomes more
interesting.  In particular in a spacetime appropriate for GPS, lacking any
exact isometries from one spacetime region to another, the ``uncertainty in
clock readings'' picks up a component from the general-relativistic curvature,
in addition to any uncertainty asserted by quantum theory.

\section{Parting Thoughts}\label{sec:7}
Experimenting with pendulums and balances in experiments done with our own hands
made us aware of a gap between a frequency ``$\omega$'' on the blackboard and
the rate of swinging of one pendulum compared to another that we could
experience on the workbench.  The key in learning to navigate between the bench
and the blackboard was to see the physical device, the paired-pendulum mechanism
of the flip-flop, that both recognizes and records a symbol.  By burying the
flip-flop under the abstract notions of spacetime and of quantum states,
theoretical physics has lost track of the rhythms and their maintenance
essential to extracting information from the bench and using that information to
control experiments.  In this report we take a first step toward bringing the
rhythms and their maintenance back into physics as a background against which
all else in physics takes place.  This background applies regardless of the mode
of explanation, and in particular regardless of whether one explains the
evidence extracted from experiments by invoking quantum theory or by invoking
general relativity.  Because quantum mechanical explanations put Planck's
constant $h$ into limits of behavior of the flip-flop, and the flip-flop works
also in the acquisition of evidence to be explained by general relativity, one
glimpses as a question for the future the a possible role of $h$ in general
relativity.

The flip-flop mimics G\"odel coding by coding whatever symbols it recognizes in
a system of numerics endowed with axioms of arithmetic. Grasping that symbols
expressed are necessarily physical prepares one to trace the influence of
physical symbols on the statements possible in physics.  So far what has been
uncovered includes the following:
\begin{enumerate}
\item Among other things, a symbol can express a pattern of other symbols, so
  that any description in physics, whether evidence or explanation, involves
  making a choice of level of detail.\vadjust{\kern-6pt}
\item Because of the separation of axioms needed to symbolize evidence from
  axioms needed to symbolize explanations, no quantum state can be determined
  from evidence without reaching beyond the evidence to exercise an irreducible
  element of free choice, \emph{i.e.} to make a \emph{guess}.\vadjust{\kern-6pt}
\item The communication of recognizable symbols requires a rhythm, and the rhythm
  requires maintenance guided not by recognitions, but by measurements
  idiosyncratic to the party making them, on which other parties in a
  communications network must rely.
\end{enumerate}

Seeing a physical mechanism for recognizing and recording a symbol opens avenues
of exploration.  Questions of ``who can know what and when can they know it?''
become colored by clock phases imposed by the pair-pendulum mechanism on which
the background of symbol exchange depends.

We offer a restructuring of \emph{clock, signal}, and \emph{time}, incorporating
attention to the recognition and recording of symbols. This structure differs
from that invoked by the IAU by bringing concepts into alignment with
practice. Recall that Einstein defined spacetime in terms of light signals
exchanged among clocks \cite{einstein16}.  We see spacetime coordinates as
implemented by devices based on the paired-pendulum mechanism for symbol
recognition (as is the implemented Turing machine). \emph{Time} amounts to
relations between the ordering by one clock of a communications network to
ordering by another clock, with the result that no isolated clock can ``tell
time.'' The ticking of clock A is influenced by the ticking of other clocks with
which clock A communicates.  Our graph pictures of evidence formalize this
structure, in which records of `digital' symbols are made in rhythms guided by
idiosyncratic `analog' measurements.

The concept of a physical basis for recognition invites application to biology.
We note that in an organism the propagation of signals goes very slowly relative
to that in electronics, so that the single oscillator that drives the clocking
throughout a digital computer likely has no biological analog; instead, we
conjecture that a nervous system, whether that of a worm or of a person,
involves rhythms in which independently adjusted oscillators take part.
Recalling the impossibility of a ``universally tightest'' communication network
in the context of general relativity, we would be interested to join other in
inquiring into constraints on coordination of such rhythms.

Questions abound concerning the role of quantum explanations in biology.  To
this topic we contribute a suggestion that DNA can be viewed as a classical code
for setting up situations, for example involving photosynthesis, describable
quantum mechanically.

 \section*{Acknowledgments}\label{sec:Acknowledgments}

 This paper grew out of a talk given at the Workshop on Topological Quantum
 Information, organized by L. H. Kauffman and S. J. Lomonaco, Jr., on May 16 and 17,
 2011, held at the Centro di Ricerca Matematica Ennio De Giorgi (CRM) in Pisa,
 Italy.  We thank the CRM for hosting the Workshop, at which several
 conversations took place that stimulated the work reported here; we also thank
 the CRM for financial help.  We thank Prof.~Kauffman for recognizing some
 condensed occurrence diagrams as virtual knot diagrams.

\end{document}